\documentclass[aps,prl,superscriptaddress,floatfix,twocolumn,a4paper,showpacs]{revtex4}
\usepackage{graphicx}
\usepackage{amsmath}
\usepackage{amssymb}
\usepackage{epsf,latexsym}
\usepackage{times}
\usepackage{epsfig}
\usepackage{citesort}

\def\be{\begin{equation}}
\def\ee{\end{equation}}
\def\bea{\begin{eqnarray}}
\def\eea{\end{eqnarray}}

\def\lsim{\raise0.3ex\hbox{$\;<$\kern-0.75em\raise-1.1ex\hbox{$\sim\;$}}}
\def\gsim{\raise0.3ex\hbox{$\;>$\kern-0.75em\raise-1.1ex\hbox{$\sim\;$}}}

\def\ie{{\it i.e.}}
\begin{document}

\title{Radiative $B-L$ symmetry breaking in supersymmetric models}

\author{S. Khalil}
\email{skhalil@bue.edu.eg} \affiliation{Centre for Theoretical
Physics, The British University in Egypt, El Sherouk City, Postal
No. 11837, P.O. Box 43, Egypt.} %
\affiliation{Department of
Mathematics, Ain Shams University, Faculty of Science, Cairo,
11566, Egypt.}
\author{A. Masiero}
\email{antonio.masiero@pd.infn.it} \affiliation{Dip. di Fisica 'G.
Galilei' and INFN, Sezione di Padova, Univ. di Padova, Via Marzolo
8, I-35131, Padua, Italy.}

\date{\today }

\begin{abstract}
We propose a scheme where the three relevant physics
 scales related to the supersymmetry, electroweak, and baryon
minus lepton ($B-L$) breakings are linked together and occur at
the TeV scale. The phenomenological implications in the Higgs and
leptonic sectors are discussed.
\end{abstract}

\maketitle

%
Nonvanishing neutrino masses and the existence of non-baryonic
dark matter (DM) represent the only two firm observational
evidences of new physics (NP) beyond the Standard Model (SM). The
energy scale(s) related to such NP are unknown with theoretical
proposals ranging from scales close to the electroweak (EW) scale
(TeV NP) to much higher scales (GUT or Planck NP). A possible
criterion to follow is to link such NP scale(s) to the breaking of
symmetries associated to the new particles appearing in the
enlarged NP particle spectrum. For instance, in the case where NP
is identified with supersymmetry (SUSY), the energy scale at which
the breaking of SUSY occurs ( in the observable sector) and the
typical mass scale for the SUSY particles have to be linked to the
(EW) scale if SUSY is called to provide the correct ultraviolet
completion of the SM to avoid the gauge hierarchy problem. In
turn, the presence of SUSY particles at the TeV scale could
provide a solution to the DM problem through the presence of the
stable lightest SUSY particle in models with the discrete symmetry
called R parity.

In the case of neutrino masses, the new particles which are
involved are likely to be the right-handed (RH) neutrinos and the
relevant symmetry to be broken should be the difference of the
baryon ($B$) and lepton ($L$) quantum numbers ($B-L$). Indeed, the
(Majorana) mass of the RH neutrino breaks $L$ or $B-L$ and, once
present, one is naturally lead to light neutrino masses through a
see-saw mechanism. However, at variance with the SUSY case, here
the breaking scale of B-L is left undetermined by the request of
obtaining a phenomenologically viable neutrino mass spectrum.

In this Letter we propose a possible link between the $B-L$ and
EW scales in SUSY models with a see-saw mechanism for neutrino
masses. Once we are in a SUSY context, we can nicely correlate the
EW and SUSY scales through the mechanism of radiative breaking
of the EW symmetry. Indeed, it
 was shown \cite{ibanez1} that radiative corrections  may drive the
 squared Higgs mass from positive
initial values at the GUT scale to negative values at the EW scale. In
such a framework, the size of the Higgs vacuum expectation value
(VEV) responsible for the EW breaking is determined by the size of
the top Yukawa coupling and of the soft SUSY breaking terms.
Analogously, we show that in a SUSY see-saw scheme it is possible
to radiatively induce the breaking of $B-L$ having the scale of
such breaking directly linked to the size of some (large) RH
neutrino Yukawa coupling and of the soft SUSY breaking scale. In
particular, we prove that for such Yukawa coupling of the order of
the top quark Yukawa coupling, the radiative mechanism leads to a
$B-L$ breaking scale of the same order as the scale of the SUSY
soft breaking terms, i.e. a TeV breaking of $B-L$.

Our result nicely fits with   a
minimal extension for the SM based on TeV scale gauge $B-L$ that has been
recently proposed \cite{Khalil:2006yi}. It was shown that this
type of models can account for current experimental results of
light neutrino masses and their large mixing \cite{Abbas:2007ag}.
In addition, the extra-gauge boson and extra-Higgs predicted in
this model have a rich phenomenology and can be detected at the
LHC \cite{Emam:2007dy}. A non-vanishing
 vacuum
expectation value (VEV), $v^\prime$, that breaks the $B-L$ gauge
symmetry was obtained in analogy with what happens for the EW
breaking. However, in such construction the scale of the scalar potential
leading to $v^\prime$ was set by hand to be of {\cal O}(1) TeV,
much in the same way that the VEV responsible for the breaking of
the EW symmetry arises from an ad hoc choice of the $\mu$ and
$\lambda$ parameters of the SM scalar potential. In this Letter we
construct a supersymmetric version of $G_{B-L}= SU(3)\times
SU(2)_L \times U(1)_Y \times U(1)_{B-L}$ model, which has been
analyzed in Ref.\cite{Khalil:2006yi,Abbas:2007ag,Emam:2007dy}. We
work out the renormalization group equation (RGE) for the relevant
parameters in the $B-L$ sector, in particular the squared mass of
the extra Higgs bosons. We study their evolution from GUT to TeV
scale and show that the squared mass of one of these Higgs bosons
can be pulled down to negative values leading to the   spontaneous
breaking of the $B-L$ symmetry. The spontaneous breaking of the
gauge $B-L$ symmetry is going to occur at a scale of {\cal O}(1)
TeV or slightly higher when the following three conditions are
met: $(i)$ The soft SUSY
 breaking terms associated to the $B-L$
sector are of order TeV. $(ii)$ The analog of the Higgs mixing
term $\mu$ in the MSSM, namely the mixing parameter of the new
Higgs superfields involved in the $B-L$ breaking, $\mu^\prime$, is
of the same size as the soft SUSY breaking terms. $(iii)$ The
Yukawa coupling of the right-handed neutrino, $Y_N = M_N/v^\prime$
is of order unity. A relevant remark is in order. In building our
extension of the MSSM,
 we introduce in the superpotential of the theory a new parameter, $\mu^\prime$,
 which
 has the dimension of a mass, in addition to the Higgs mixing $\mu$ parameter
 of the MSSM. As known,
this latter parameter is present in the SUSY invariant part of the
theory and hence its scale is not directly set by the scale of the
soft breaking parameters. Why $\mu$ should then be at the TeV
scale and not, for instance, at the superlarge scale of
supergravity breaking constitutes the so-called $\mu$ problem. A
possible suggestion to obtain a $\mu$ scale of the order of the EW
scale is known as the Giudice-Masiero mechanism
\cite{Giudice:1988yz}. Here we are advocating that this same
mechanism could be responsible also for the origin of the
$\mu^\prime$ parameter, hence implying a similar mass scale for
both of them.

We consider the minimal supersymmetric version of
the $B-L$ extension of the SM based on the gauge group
$G_{B-L}=SU(3)_C \times SU(2)_L \times U(1)_Y \times U(1)_{B-L}$.
 The particle content of the SUSY $B-L$ includes the following fields
 in addition to those of the MSSM: three
chiral right-handed superfields ($N_i$), the vector superfield
necessary to gauge the $U(1)_{B-L}$ ($Z_{B-L}$), and two chiral
SM-singlet Higgs superfields ($\chi_1$, $\chi_2$ with $B-L$
charges $Y_{B-L}=-2$ and $Y_{B-L}=+2$, respectively). As in the MSSM,
the introduction of a second Higgs singlet ($\chi_2$) is necessary
in order to cancel the $U(1)_{B-L}$ anomalies produced by the
fermionic member of the first Higgs ($\chi_1$) superfield. The
$Y_{B-L}$ for quark and lepton superfields are assigned in the
usual way.

 The interactions between Higgs and matter superfields are
described by the superpotential %
\begin{eqnarray}%
W &=& (Y_U)_{ij} Q_i  H_2 U^c_j + (Y_D)_{ij} Q_i  H_1 D_j^c +
(Y_L)_{ij} L_i H_1 E_j^c  \nonumber\\
&+& (Y_\nu)_{ij} L_i H_2 N_j^c + (Y_N)_{ij}
N^c_i N^c_j \chi_1 + \mu H_1 H_2 \nonumber\\
&+& \mu^\prime \chi_1 \chi_2.%
\label{superpot}%
\end{eqnarray}

Interestingly enough, the presence of the $B-L$ gauge symmetry,
forbids the appearance in  the superpotential of the $B$ or $L$
violating terms. Hence, in this model there is no need to impose
an additional discrete symmetry ( for instance, $R$ parity) to
achieve such result.

Assuming flavor and gaugino universality at the grand unification scale,
$M_X = 3 \times 10^{16}$ GeV,
the SUSY soft breaking Lagrangian at that scale reads
\begin{eqnarray}
- {\cal L}_{soft} &=&m_0^{2}\left[\vert{\widetilde{Q_i}}\vert^2
+\vert{\widetilde{U_i}}\vert^2  + \vert{\widetilde{D_i}}\vert^2
+\vert{\widetilde{L_i}}\vert^2 +\vert{\widetilde{E_i}}\vert^2 \right.\nonumber\\
&+& \left. \vert{\widetilde{N_i}}\vert^2 + \vert{H_1}\vert^2 +
\vert{H_2}\vert^2 + \vert{\chi_1}\vert^2
+\vert{\chi_2}\vert^2\right]\nonumber\\
&+& \left[Y_{U}^{A}{\widetilde{Q}} {\widetilde{U}}^{c} H_{2} +
Y_{D}^{A}{\widetilde{Q}} {\widetilde{D}}^{c} H_{1} +
Y_{E}^{A}{\widetilde{L}} {\widetilde{E}}^{c} H_{1}
\right.\nonumber\\
&+& \left. Y_{\nu}^{A}{\widetilde{L}} {\widetilde{N}}^{c} H_{2} +
Y_{N}^{A}{\widetilde{N}}^{c}
{\widetilde{N}}^{c}\chi_{1} + B (\mu H_1 H_2 \right.\\
&+& \left. \mu^\prime \chi_1 \chi_2) + h.c\right] + \frac{1}{2}
M_{1/2}\left[{\widetilde{g}}^a {\widetilde{g}}^a +
{\widetilde{W}}^a {\widetilde{W}}^a \right. \nonumber\\
&+& \left.{\widetilde{B}}{\widetilde{B}} +
{\widetilde{Z}_{B-L}}{\widetilde{Z}_{B-L}}+ h.c \right] \;,\nonumber%
\label{Lsoft}%
\end{eqnarray}%
where $(Y^A)_{ij}\equiv (YA)_{ij}$. The tilde denotes the scalar
components of the chiral matter superfields and fermionic components
of vector superfields. We denote by $H_{1,2}$ and $\chi_{1,2}$
also the scalar components of the Higgs superfields $H_{1,2}$ and
$\chi_{1,2}$.

Let us now discuss how the $B-L$ and electroweak symmetries may be
broken in the SUSY $G_{B-L}$. We have to study the scalar
potential for the Higgs fields $\chi_{1,2}$ and $H_{1,2}$ and
check if there are minima for which $\langle \chi_1\rangle,
\langle \chi_2\rangle \neq 0 $ and $\langle H_1\rangle, \langle
H_2\rangle \neq 0 $.
The scalar potential for $H_{1,2}$ and $\chi_{1,2}$ is
\begin{eqnarray}
V(H_1,H_2,\chi_1,\chi_2) & = & \frac{1}{2}
g^2(H_1^*\frac{\tau^a}{2} H_1 + H_2\frac{\tau^a} {2}H_2)^2
\nonumber\\
&&\hspace{-1.5cm}+~\frac{1}{8} g^{\prime^2}(\vert H_2\vert^2 -
\vert H_1\vert^2)^2 + \frac{1}{2} g^{\prime\prime^2}(\vert
\chi_2\vert^2 - \vert \chi_1\vert^2)^2
\nonumber\\
&& \hspace{-1.5cm}+~ m_1^2 \vert H_1\vert^2 +
            m_2^2\vert H_2\vert^2 - m_3^2(H_1 H_2 + h.c)
            \nonumber\\
            && \hspace{-1.5cm}+~\mu^2_1 \vert \chi_1\vert^2 +
            \mu^2_2\vert \chi_2\vert^2 - \mu^2_3(\chi_1 \chi_2 + h.c)  ,
\label{Vtotal}
\end{eqnarray}
where
\begin{eqnarray}
m^2_i &=& m_0^2 + \mu^2 , \hspace{0.25cm} i=1,2 \hspace{1cm} m_3^2
=- B \mu~, \label{m12}\\
\mu^2_i &=& m_0^2 + \mu^{\prime^2} , \hspace{0.2cm} i=1,2
\hspace{1cm} \mu^2_3 = - B \mu^\prime~.%
\label{mp12}%
\end{eqnarray}
As can be seen from Eq.(\ref{Vtotal}), the potential
$V(H_1,H_2,\chi_1,\chi_2)$ results from the sum of the usual
 MSSM scalar potential  $V(H_1,H_2)$ and of the new potential
 $V(\chi_1,\chi_2)$ which exhibits an apparent similarity in its structure
 to $V(H_1,H_2)$. For simplicity, in defining $\mu^2_3$ and $m_3^2$ only one
$B$ parameter has been introduced.

As is known, the radiative breaking of the EW symmetry
is induced by the running
from  $M_X$ to the weak scale of $m_2^2$. Given the large value
of the top Yukawa coupling, such running succeeds to turn the positive
value of $m_2^2$ at $M_X$ to a negative value, hence inducing the desired
EW breaking.

Following the same steps as for the minimization of $V(H_1,H_2)$ in the MSSM,
one readily obtains for the minimization of $V(\chi_1,\chi_2)$:
\be %
v^{\prime^2} = (v^{\prime^2}_1 + v^{\prime^2}_2) =
\frac{(\mu^2_1-\mu^2_2) - (\mu^2_1+\mu^2_2)\cos2\theta}{2
g^{\prime\prime^2}
\cos2\theta},%
\label{muprime}
\ee %
where $\langle \chi_1 \rangle = v^{\prime}_1$ and $\langle \chi_2
\rangle = v^{\prime}_2$. The angle $\theta$ is defined as $\tan
\theta = v^{\prime}_1/v^{\prime}_2$. Consequently,
the $Z_{B-L}$ gauge boson  acquires a mass
\cite{Khalil:2006yi}: $M^2_{Z_{B-L}}=4 g^{\prime\prime^2}
v^{\prime^2}$.

The boundness from below of the potential $V(\chi_1,\chi_2)$ requires
\begin{equation}
\mu_1^2 + \mu_2^2 > 2 \vert \mu_3^2\vert. %
\label{stab}
\end{equation}
This represents the stability condition for the potential. Furthermore, to avoid
that $\langle \chi_1 \rangle = \langle \chi_2 \rangle  = 0$ be a
local minimum one has to impose that
\begin{equation}
\mu_1^2 \mu_2^2 < \mu_3^4. \label{minimiz}
\end{equation}
It is not possible to simultaneously fulfill both the above
conditions for the positive values of $\mu_1^2$ and $\mu_2^2$ as
given in Eq.(\ref{m12}). Indeed, if $ \mu_1^4 = \mu_2^4 < \mu_3^4
$, then the condition Eq. (\ref{stab}) is not satisfied and the
scalar potential is unbounded from below in the direction $
\langle \chi_1 \rangle = \langle \chi_2 \rangle \rightarrow
\infty$.

The problem we encounter is reminiscent of what occurs for the
electroweak symmetry breaking, i.e. for the $V(H_1,H_2)$ part of
 the potential $V(H_1,H_2,\chi_1,\chi_2)$. In that case,
 the problem of obtaining the desidered breaking vacuum while guaranteeing
the stability of the potential is solved~\cite{ibanez1} by noting
 that the boundary conditions
Eq.(\ref{m12}) are valid only at the GUT scale. However, in the
running from that large scale down to $M_W$, one finds that $m^2_1$
and $m^2_2$ get renormalized differently if $H_1$ and $H_2$ couple
with different strength to fermions. Indeed, $H_2$ couples to the
top quark with a large Yukawa coupling. The running from $M_X$
down to the weak scale reduces the squared Higgs masses, until
eventually the minimization condition is satisfied and the
electroweak gauge symmetry is broken.

We consider the $B-L$ renormalization group equations and analyze
the running of the scalar masses $m_{\chi_1}^2$ and
$m_{\chi_2}^2$. The key point for implementing the radiative $B-L$
symmetry breaking is that the scalar potential $V(\chi_1,\chi_2)$
receives substantial radiative corrections. In particular, a
negative (mass)$^2$ would trigger the $B-L$ symmetry breaking of
$B-L$. We argue that the masses of Higgs singlets $\chi_1$ and
$\chi_2$ run differently in the way that $m^2_{\chi_1}$ can be
negative whereas $m^2_{\chi_2}$ remains positive. The
renormalization group equation (RGE) for the $B-L$ couplings and
mass parameters can be derived from the general results for SUSY
RGEs of Ref.\cite{Martin:1993zk}. Here, for simplicity, we neglect
the couplings of the first two generations. As is known,
neglecting the Yukawa couplings of the first two generations for
the SM quark and lepton is quite justified approximation due to
the smallness of their masses. However, for the Yukawa coupling
$h_N$, this is a further assumption. Also it is more convenient to
write the RGE in terms of gauge couplings: $\tilde{\alpha}_i =
g_i^2/16 \pi^2$ and Yukawa couplings: $\tilde{Y}_i=
Y^2_i/16\pi^2$.

The RGEs for the masses of the $B-L$ Higgs
field $\chi_1$ and right-handed sneutrino read
\bea %
\!\!\frac{d m^2_{\chi_1}}{d t} \!&\!=\!&\! 6 \tilde{\alpha}_{B-L}
M_{B-L}^2 \!-\!
2\tilde{Y}_{N_3} \left(m^2_{\chi_1}+ 2 m^2_{N_3} + A_{N_3}^2 \right)\!,~~~~\\
\!\!\frac{d m^2_{{N_3}}}{d t} \!&\!=\!&\! \frac{3}{2}
\tilde{\alpha}_{B-L} M_{B-L}^2 \!-\!
\tilde{Y}_{N_3} \left(m^2_{\chi_1}+ 2 m^2_{N_3} + A_{N_3}^2 \right)\!.~~~~ %
\eea %

Since the second Higgs $\chi_2$ has no interaction with any
particle, its evolution is given by%
\be %
\frac{d m^2_{\chi_2}}{d t} = 6 \tilde{\alpha}_{B-L} M_{B-L}^2. %
\ee %
The evolution of these mass parameters depends on the boundary
conditions at GUT scale. As mentioned, we assume universal soft
SUSY breaking at this scale, \ie, %
\bea %
m^2_{\chi_1}(0) &=& m^2_{\chi_2}(0) = m^2_{N_3}(0)= m_0^2,\\
M_a(0) &=& M_{B-L} = M_{1/2}, ~~\nonumber\\
a&=&1,2,3~ \mathrm{for}~ SU(3)_C,SU(2)_L,U(1)_Y\\
A_i(0) &=& A_{N_3} = A_0, ~~~~~~~~~~~~~~ i=t,b,\tau.%
\eea

\begin{figure}[t]
\epsfig{file=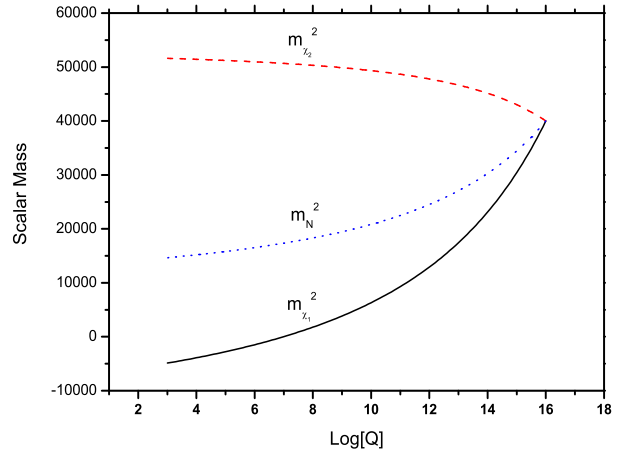,width=9cm,height=7.cm}%
\caption{The evolution of the $B-L$ scalar masses from GUT to TeV
scale for $m_0 = M_{1/2}= A_0=200$ GeV and $Y_{N_3}\sim {\cal
O}(0.1)$. }
\label{mass-running}
\end{figure}
%
Fig. \ref{mass-running} reports the result of the running. In this
figure, we set $m_0 = M_{1/2}= A_0=200$ GeV and order one $Y_{N_3}
\simeq M_{N_3}/v'$ is assumed. As can be seen from this figure,
$m^2_{\chi_1}$ drops rapidly to negative region, while
$m^2_{\chi_2}$ remains positive. Analogously to the radiative
electroweak symmetry breaking, this mechanism works for large
Yukawa coupling. It is worth noting  the faster drop of
$m^2_{\chi_1}$ in comparison with that of $m^2_{H_2}$. Indeed,
 $m^2_{\chi_1}$
receives a positive contribution in its running only from the
$B-L$ gaugino, while the $SU(2)_L$ and $U(1)_Y$ gaugino masses
are responsible for the positive contributions in the running of
 $m^2_{H_2}$.

Also in Fig. \ref{mass-running}, we plot the scale evolution for
the scalar mass $m^2_{N_3}$. Although  $m^2_{N_3}$ decreases in
 the running from $M_X$, it remains positive at the TeV scale.
Therefore, the $B-L$ breaking via a non-vanishing vacuum
expectation value for right-handed sneutrino does not occur in the
present framework.

The phenomenology of TeV scale neutral gauge boson $Z_{B-L}$ is
very rich and its potential discovery at LHC has been recently
analyzed in Ref.\cite{Emam:2007dy}. Also, the three SM singlet
fermions, $\nu_{R_i}$ in the superfields
$N_i$, get the following masses: %
\be %
M_{N_i}= v'  Y_{N_i} \sim {\cal O}(\rm{TeV}). %
\ee %
These three particles play the role of right handed neutrinos. In
addition, the electroweak symmetry breaking induces the Dirac
mass term:%
\be%
m_D = \frac{v}{\sqrt{2}} Y_{\nu}.%
\ee%
Therefore, the observed light-neutrino masses can be obtained
through the usual seesaw mechanism with Yukawa neutrino coupling,
$Y_{\nu}$, of order ${\cal O}(10^{-6})$ \cite{Abbas:2007ag}.

The Higgs sector of this model consists of two Higgs doublets and
two Higgs singlets with no mixing. However, after the $B-L$
symmetry breaking, one of the four degrees of freedom contained in
the two complex singlet $\chi_1$ and $\chi_2$ is swallowed in the
usual way by the $Z^0_{B-L}$ to become massive. Therefore, in
addition to the usual five MSSM Higgs bosons, namely one neutral
pseudoscalar Higgs bosons $A$, two neutral scalars $h$ and $H$ and
a charged Higgs boson $H^{\pm}$,  three new physical degrees of
freedom remain. They form a neutral pseudoscalar Higgs boson $A'$
and two neutral scalars $h'$ and $H'$. Their masses at tree level
are given by %
\begin{equation}
m_{A'}^2 = \mu_1^2 + \mu_2^2,
\end{equation}
\begin{eqnarray}%
m_{H',h'}^2 &=& \frac{1}{2}\left(m_A'^2 + M^2_{Z_{B-L}} \right.
\nonumber\\
&\pm& \left. \sqrt{(m_{A'}^2 + M_{Z_{B-L}}^2)^2-4m_A^{'^2}
M_{Z_{B-L}}^2\cos 2 \theta}\right).~~~~~~
\end{eqnarray}
Here $\theta= \tan^{-1} \frac{v'_1}{v'_2}$ and $\mu_{i}$ with
$i=1,2$ are defined in Eq.(\ref{mp12}). From the expression of the
lightest $B-L$ Higgs boson, one finds the following upper bound
\begin{equation}
m_h'  \leq M_{Z_{B-L}} \vert \cos 2 \theta \vert .
\end{equation}
However, in analogy with the large radiative corrections to the
lightest MSSM Higgs mass due to the top-stop loop, the
$N-\tilde{N}$ loop  can induce  large correction leading to
 $m_{h'} > m_{Z'}$.

The enlarged sneutrino sector of this model deserves some attention.
 Indeed, in the present SUSY extension of the
$G_{B-L}$ model, a significant mixing between the left-handed and
right-handed sneutrinos can be obtained. This would lead to what
is known as sneutrino-antisneurino oscillation
\cite{Dedes:2007ef}. The $12 \times 12$ sneutrino mass matrix, in
the basis $(\phi_L, \phi_N)$ with $\phi_L = (\tilde{\nu}_L,
\tilde{\nu}_L^*)$ and $\phi_{N}=(
\tilde{\nu}_R, \tilde{\nu}_R^*)$, is given by %
\be %
M^2 = \frac{1}{2} \left(\begin{array}{cc}
                M^2_{LL} & M^2_{LN}\\
                M^2_{NL} & M^2_{NN}
                \end{array}\right).
\ee %
The detailed expressions for the $6\times 6$ matrices $M^2_{AB}$,
for $A,B = L,N$ can be found in Ref. \cite{Dedes:2007ef}. In
general, the order of magnitude of the entries
of this matrix  can be estimated as follows:%
\be %
M^2 = \frac{1}{2} \left(\begin{array}{cc}
                {\cal O}(v^2) & {\cal O}(v v')\\
                {\cal O}(v v') & {\cal O}(v^{'^2})
                \end{array}\right).
\ee %
Since $v'\sim $ TeV, the sneutrino matrix elements are of the same
order and there is no seesaw type behavior as usually found in
MSSM extended with heavy right-handed neutrinos. Therefore a
significant mixing among the left- and right- handed sneutrinos is
obtained. The phenomenological consequences for such mixing have
been studied in \cite{Grossman:1997is}.\\

In conclusion, we have shown that in a SUSY extension of the SM
where $B-L$ is gauged, it is possible to link together  the
electroweak, $B-L$ and soft SUSY breakings at a scale of O(TeV).
The ensuing richer TeV phenomenology   for the coming LHC
 and  neutrino physics opens new prospects and deserves further
 attention.\\

\section*{Acknolegements}
One of us (A.M.) ackowledges the support of the European Research
Training Network HPRN-CT-2006-035863 ( UniverseNet) and of the
PRIN 2006023491 ( "Fundamental Constituents of the Universe") of
the Italian Ministry of University and Research. The work of S.K.
was partially supported by the ICTP through the OEA-project-30.



\begin{thebibliography}{99}
%

\bibitem{ibanez1}
L.~E.~Ibanez and G.~G.~Ross,
  Phys.\ Lett.\  B {\bf 110}, 215 (1982);
  K.~Inoue, A.~Kakuto, H.~Komatsu and S.~Takeshita,
  Prog.\ Theor.\ Phys.\  {\bf 68}, 927 (1982)
  [Erratum-ibid.\  {\bf 70}, 330 (1983)];
  L.~Alvarez-Gaume, J.~Polchinski and M.~B.~Wise,
  Nucl.\ Phys.\  B {\bf 221}, 495 (1983);
  L.~E.~Ibanez and G.~G.~Ross,
  arXiv:hep-ph/9204201.
\bibitem{Khalil:2006yi}
  S.~Khalil,
  arXiv:hep-ph/0611205.
\bibitem{Abbas:2007ag}
  M.~Abbas and S.~Khalil,
  arXiv:0707.0841 [hep-ph].
\bibitem{Emam:2007dy}
  W.~Emam and S.~Khalil,
  arXiv:0704.1395 [hep-ph], to appear in Eur.Phys.J.C.
\bibitem{Giudice:1988yz}
  G.~F.~Giudice and A.~Masiero,
  Phys.\ Lett.\  B {\bf 206}, 480 (1988).
\bibitem{Martin:1993zk}
  S.~P.~Martin and M.~T.~Vaughn,
  Phys.\ Rev.\  D {\bf 50}, 2282 (1994)
  [arXiv:hep-ph/9311340]. See also,
  K.~Inoue, A.~Kakuto, H.~Komatsu and S.~Takeshita,
  Prog.\ Theor.\ Phys.\  {\bf 68}, 927 (1982)
  [Erratum-ibid.\  {\bf 70}, 330 (1983)]; J.~R.~Ellis, J.~S.~Hagelin, D.~V.~Nanopoulos and K.~Tamvakis,
  Phys.\ Lett.\  B {\bf 125}, 275 (1983);  L.~E.~Ibanez and C.~Lopez,
  Phys.\ Lett.\  B {\bf 126}, 54 (1983);   L.~Alvarez-Gaume, J.~Polchinski and M.~B.~Wise,
  Nucl.\ Phys.\  B {\bf 221}, 495 (1983).
\bibitem{Dedes:2007ef}
  A.~Dedes, H.~E.~Haber and J.~Rosiek,
  arXiv:0707.3718 [hep-ph].
\bibitem{Grossman:1997is}
  Y.~Grossman and H.~E.~Haber,
  Phys.\ Rev.\ Lett.\  {\bf 78}, 3438 (1997)
  [arXiv:hep-ph/9702421];
\end{thebibliography}
\end{document}